\documentclass[a4paper]{article}

\usepackage{INTERSPEECH2022}
\usepackage{indentfirst} 
\usepackage{amssymb}
\usepackage{pifont}
\usepackage{algorithm}
\usepackage[noend]{algpseudocode}
\usepackage{footmisc}

\title{Karaoker: Alignment-free singing voice synthesis with speech training data}
\name{
	\begin{tabular}{c}
		Panos Kakoulidis$^{\star}$,
		Nikolaos Ellinas$^{\star}$,
		Georgios Vamvoukakis$^{\star}$,
		Konstantinos Markopoulos$^{\star}$, \\
		June Sig Sung$^{\dagger}$,
		Gunu Jho$^{\dagger}$,
		Pirros Tsiakoulis$^{\star}$,
		Aimilios Chalamandaris$^{\star}$
	\end{tabular}
}
\address{$^{\star}$ Innoetics, Samsung Electronics, Greece \\
	$^{\dagger}$ Mobile Communications Business, Samsung Electronics, Republic of Korea}
\email{\{p.kakoulidis, n.ellinas, g.vamvouk, k.markop,\
	js6.sung, gunu.jho, p.tsiakoulis, aimilios.ch\}@samsung.com }

\begin{document}

\maketitle
\begin{abstract}
Existing singing voice synthesis models (SVS)
are usually trained on singing data and
depend on either error-prone time-alignment and duration features or explicit music score information. In this paper, we 
propose  Karaoker,  a  multispeaker 
Tacotron-based model conditioned  on  voice  characteristic features that is trained exclusively on spoken data without requiring time-alignments. 
Karaoker  synthesizes  singing voice and transfers style following  a  multi-dimensional 
template extracted from a source waveform of an unseen 
singer/speaker.  The  model  is  jointly  conditioned  with  a 
single  deep convolutional encoder  on  continuous  data  including 
pitch,  intensity,  harmonicity,  formants,  cepstral  peak 
prominence  and  octaves.  We  extend  the  text-to-speech 
training  objective  with feature reconstruction, classification and speaker identification tasks that guide the model to an accurate result.  In addition to multi-tasking, we also employ a Wasserstein GAN training scheme as well as new losses on the acoustic model's output to further refine the quality of the model.
\end{abstract}
\noindent\textbf{Index Terms}: singing voice synthesis, voice style transfer, prosody transplantation, sequence-to-sequence, conditioning, multi-tasking, Tacotron, Wasserstein GAN, convolutional

\section{Introduction}

During the recent years, singing voice synthesis (SVS) is following the progress on speech synthesis via the introduction of deep neural architectures. Synthesizing singing vocals is a special
case of human voice generation and more complex than classical text-to-speech (TTS). This kind of synthesis aims on producing a highly variable and accurate result that complies with a specific melody from a source waveform (template-based SVS) or a music score (score-based SVS). The generated vocals should fulfill the requirement of naturalness and preserve the identity of the target voice, regardless of any imposed high variations in prosody. Ongoing research has proposed several deep learning solutions that generate waveforms in an end-to-end manner or in two separate stages; acoustic modeling and vocoding. \par
However, many existing SVS methods require resources that are either expensive or hard to obtain. These models utilize singing training data, which are more challenging to create due to the overhead work (transcribing or collecting information on target melodies, possible re-takes for interpretation of the reference, selection/employment of professional singers) and higher costs. Some models rely on data from automatic lyric-to-audio alignments, which burden the feature preparation and contain errors that affect the resulting quality of the model. Also, score-based SVS models demand music transcriptions which are highly expensive resources.

\subsection{Related Work}
 Until recently, the majority of the state-of-the-art models are score-based SVS approaches that are trained exclusively on singing data. These methodologies leverage the work on neural TTS models with prosody control capabilities. Several speech synthesis neural architectures have been adapted to the SVS task to meet its high demand on melodic accuracy. For instance, there are proposals that re-purpose \cite{Angelini2020} or extend Tacotron--an attention-based sequence-to-sequence model \cite{8461368}--with duration and pitch encoders for conditioning \cite{wang2022singing}, while others have also incorporated additional modules like a duration model \cite{gu2021bytesing} or a discriminator \cite{Lee2019}. \par 
 Singing voice synthesis via deep neural models with spoken-only training data  has not been explored thoroughly, except for some recent attempts during the last years. One of the first studies on this topic was Mellotron \cite{9054556}, a multispeaker GST-Tacotron 2 model conditioned on pitch contours that focuses on expressive text-to-speech and inspired the creation of our model. It requires text-to-audio alignments as input which are obtained either by forced-alignment or by manually finetuned attention weights on the source waveform that contains the style to be transferred. A Cross-lingual SVS  model that can be trained on a bilingual speech dataset and generate vocals in any of the two languages is also presented in \cite{9362077}. M-U model \cite{choi2021melody} offers an alternative option for finetuning on speech data but input pitch values are quantized allowing limited control and the vocoder is trained on singing data. Our previous work \cite{markopoulos21_ssw} explores singing-data-free training by combining a TTS prosody control model \cite{10.1007/978-3-030-87802-3_11} with a post-processing DSP module, resulting to a melodic voice generation of high quality but with limited pitch variation.
 
\subsection{Proposed method}
A fundamental SVS problem is to produce a precise flow of vocals according to the timescale dictated by a music sheet. Aligning the spoken phonemes to their corresponding waveform is an active research topic for free style speech synthesis. Singing adds to this complexity with a lot more different voicing variations for the same phoneme like the vibrato technique. As we already mentioned, some previous approaches relax the problem of alignment by learning estimations of phoneme durations given a phoneme/note sequence with supervised duration modeling. However, this method still depends on offline lyrics-to-audio alignments for training data, either parametrically generated with a considerable error rate or curated by experts similarly to music scores. Our proposal sticks to an unsupervised end-to-end learning approach that circumvents the need of such resources.

In this paper, we focus on improving the base multispeaker Tacotron 2 acoustic model for a templated-based singing voice synthesis in a constrained setting that lacks domain-specific resources: singing training data, music scores, time-alignments of input features, limited data availability per speaker.  This is achieved through the following contributions:

\begin{itemize}
	\item The acoustic model is conditioned on multiple locations with an embedding that represents a feature set on voice characteristics extracted from a source audio.
	\item The quality of the generated voice is further enhanced by incorporating prior work on the refinement of attention modeling \cite{tachibana2018efficiently, Ellinas2020, kang2021fast} and by inserting multiple-tasks, additional losses and a discrimination phase into the training objective.

\end{itemize}

 \noindent The proposed acoustic model is able to transfer the style of a voice that belongs to an unseen speaker or singer. The gender of the speakers in the training set is taken into account since the multi-speaker model compensates for any inadequacies in pitch coverage with data from other speakers in the training data.

\section{Method}

\subsection{Feature selection}

Since there are no singing data involved in the training, a number of hooks in the acoustic model are needed for a transition path towards a melodic voice style. We model the human voice on the dimensions specified by unsupervised feature selection  \cite{cai2010unsupervised} between several features including ones that are pertinent to phonation mode classification in singing \cite{stoller-2016-analysis}. The selection process was made on an internal speech dataset and the frame-level features selected were: pitch (F0), harmonicity (HNR), cepstral peak prominence (CPP), intensity (RMS), formants (F1-F4), octave.

\subsection{Acoustic model architecture}

\begin{figure}[t]
	\centering
	\includegraphics[width=\linewidth]{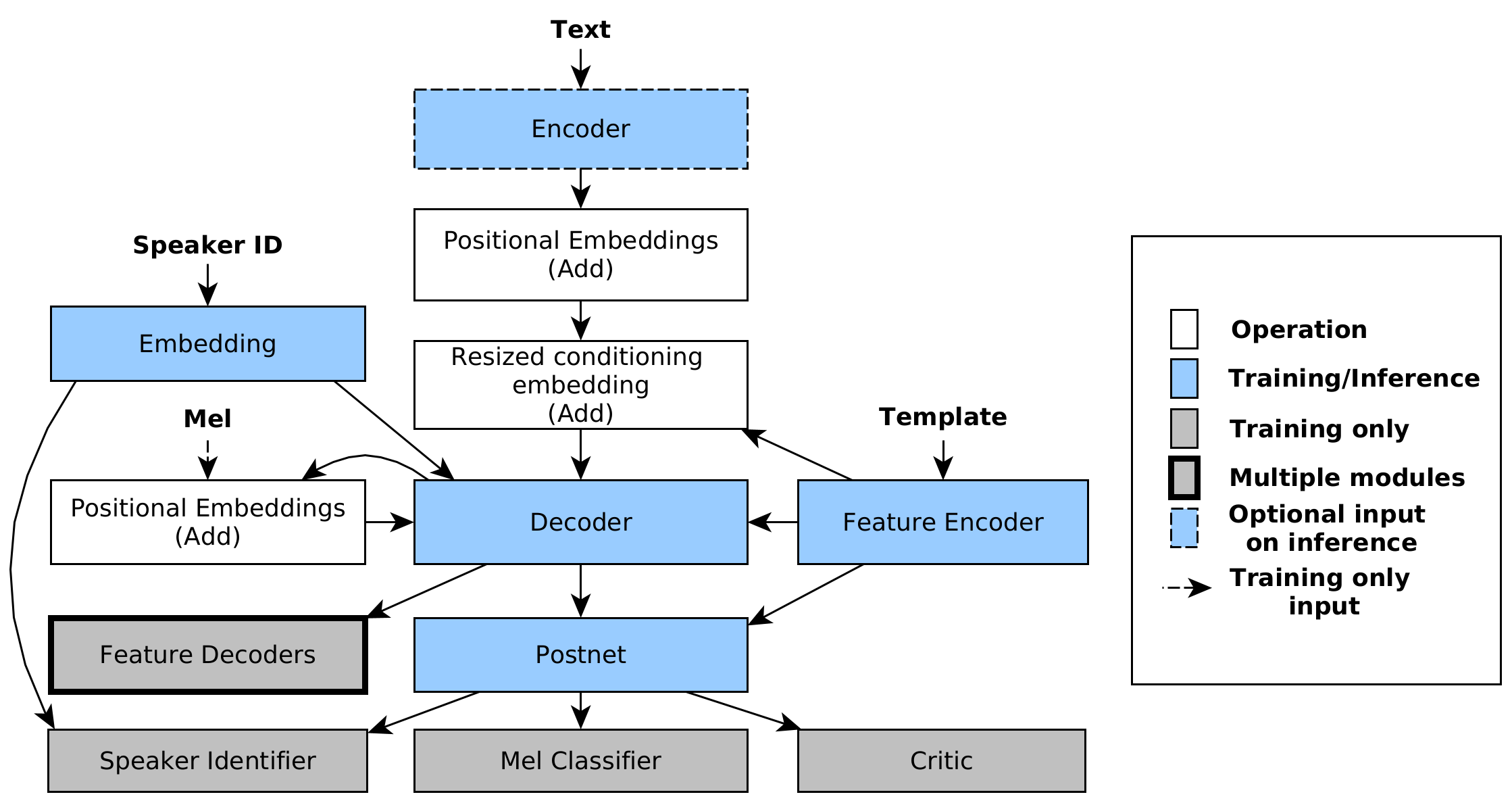}
	\caption{Karaoker's architecture}
	\label{fig:architecture}
\end{figure}

The architecture of the proposed acoustic model (Figure \ref{fig:architecture}) is based on Tacotron 2, a sequence-to-sequence model that aligns phoneme sequence embeddings to mel spectrogram frames with an attention mechanism and decodes these encoded representations to mel spectrograms. We replace Location-Sensitive-Attention (LSA) with  Mixture-of-Logistics (MoL) attention as described in \cite{Ellinas2020}. We add positional embeddings on the inputs of attention model \cite{kang2021fast} and we adopt Guided Attention loss \cite{tachibana2018efficiently} as a means to improve the alignment.

A feature encoder conditions the acoustic model with a series of 8-channeled embeddings extracted from our selected feature set and masked according to their padding. The encoder consists of 8 blocks; each block contains 3 stacked 1x1 convolutional layers (channels: [8,8], [8,64], [64,8]) with batch normalization and exponential linear unit (ELU) activations. We concatenate the feature encoder outputs with the phoneme encoder outputs after performing nearest neighbors interpolation to match their lengths. During each step of the decoder, we condition the Prenet by concatenating a projection of a conditioning embeddings' slice that corresponds to the current step's frames. The attention mechanism also receives this projection along with a projected mean of the embeddings along the time axis. Finally, the projected conditioning embeddings are added to the decoder outputs and given as input to the post-net.

The acoustic model is trained as a generator under a Wasserstein GAN training scheme \cite{wei2018improving} to further improve its quality. We built a Critic with 4 uDisLayers \cite{Wu2020}, which is jointly trained with the acoustic model, discriminating randomly drawn samples from random locations of the ground truth and the generated output. The Critic's training starts when the acoustic model reaches a fair accuracy. The update of the acoustic model by the Critic begins when the discrimination outputs stably diverge for real and fake inputs.

\subsection{Multi-tasking} 

To the best of our knowledge, multi-task learning has not been investigated before in SVS. By attaching multi-tasking sub-networks during training, we managed to ameliorate the acoustic model without modifications.  We emphasize on the naturalness, the precision on music notes and the preservation of speaker identity, thus we formulated the following auxiliary tasks with these principles in mind:

\begin{itemize}
	\item Real/fake classification of random-sized windows from random positions from mel spectrograms in training data and post-net outputs. The module includes 2 1x1 convolutions with ELU activation  (channels: [80,80] , [80,1]) and a  mean sigmoid output.
	\item Reconstruction of F0, HNR, RMS features from decoder outputs. For each feature, there is a dedicated feature decoder that consists of 3 1x1 convolutional layers with ELU activations (channels: [80,80], [80,32], [32, 1]). The values and their rate of change are evaluated separately with the version of SoftDTW from \cite{elias2021parallel}.
	\item Speaker identification on the generated mel spectrogram. The module contains 4 1x1 convolutions (channels: [80,80] , [80,64], [64,32], [32, 64]) with ELU activations and max pooling (only the first 2 layers, dimension: 3).  A similar method is also proposed in \cite{Cai2020} but the speaker embedding layer is frozen during the verification step. 
\end{itemize}

\subsection{Losses} 

We integrate two new criteria in the objective function of the acoustic model's training that validate the predicted mel spectrograms: normalized mel rate and decomposition losses. The first loss evaluates the accuracy of notes sung in the same manner for all samples in the batch, regardless of the underlying scale (key) of each music piece. The closest reference we found on this suggestion is in \cite{Shechtman2019}, which is a harder constraint hurting generalization.

\begin{align}
	\begin{split}
	L_{mr} = \big\Vert f_{rate}(mel) - f_{rate}(mel_{dec})\big\Vert_{1} + \\ 
	\big\Vert f_{rate}(mel) - f_{rate}(mel_{post})\big\Vert_{1}
    \end{split}
	\label{eq1}
\end{align}

\noindent where $mel$ is the ground truth,  $mel_{dec}$, $mel_{post}$ refer to the decoder and post-net outputs. $f_{rate}$ is defined as:
\begin{equation}
	f_{rate}(\hat{x}) =f_{scale}(x_{t} - x_{t-1} \,\, \forall  \,\, 1 \leq t \leq T)
	\label{eq2}
\end{equation} 

\noindent where $t$ is the timestep, $T$ is the total number of mel frames and $f_{scale}$ is:
\begin{equation}
	f_{scale}(\hat{x}) = exp\left(\frac{x_{m} - min(\hat{x})}{max(\hat{x}) - min(\hat{x})}\right) \, \forall  \,\, 0 \leq m \leq M
	\label{eq3}
\end{equation} 

\noindent where $M$ is the length of $\hat{x}$. Also, we propose a loss on the difference between the frequency basis vectors of ground truth mel spectrogram and postnet outputs. We retrieve these vectors through the method of Singular Value Decomposition (SVD) and we formulate the decomposition loss as:

\begin{equation}
	L_{svd} = \big\Vert U_{mel} - U_{mel\,post}\big\Vert_{1}
	\label{eq4}
\end{equation}

\noindent where $U$ is a basis vector stemming from the factorization of a $mxn$ matrix $A$ (in this context, the mel spectrogram) via SVD:
\begin{equation}
 A = U \Sigma V^{T}
 \label{eq5}
\end{equation}

\noindent Regarding the multi-tasking losses, $L_{rec}$ is the reconstruction loss  between the feature ($feat_{i}$) and the feature decoding ($dec_{i}$). $feat_{i}$ has frame-level resolution and is padded to mel length. This loss is computed for $N$ features of $L$ length and is defined with the help of (\ref{eq2}) as:
  	\begin{multline}
 		L_{rec} =  \sum_{i=0}^{N}\left( \frac{1}{L} \sum_{j=0}^{L}\left(softdtw(feat_{i}, dec_{i})\right) + \right. \\
 		\left. \frac{1}{L} \sum_{j=0}^{L}\left( softdtw(f_{rate}(feat_{i}), f_{rate}(dec_{i})) \right) \vphantom{\sum_{j=0}^{L}} \right)
		\label{eq6} 
	\end{multline}

\noindent The classification loss is computed as:
  \begin{equation} 
	L_{class} = L_{real} - L_{fake}
	\label{eq7} 
\end{equation}
\noindent and the equation for the speaker identification loss is :
\begin{equation} 
	L_{spk} = 1 - cos(x_{spk}, P(x_{post}))
	\label{eq8} 
\end{equation}
\noindent where $x_{spk}$ is a speaker embedding, $x_{post}$ is a post-net output and $P$ is a projection operator. Therefore, the total loss function is formed as:
\begin{equation}
	\begin{split}
		L = L_{mel} + L_{gate} +  L_{svd} + L_{mr} +  L_{att} \\  + L_{rec} + L_{class} + L_{spk} - L _{D}
		\label{eq9} 
	\end{split}
\end{equation}

\noindent where $L_{mel}$ are the $L_{1}$ losses for [mel, decoder outputs] and [mel, post-net outputs] pairs, $L_{gate}$ is the prediction loss for the stop token, $L_{att}$ is the Guided Attention loss (following the implementation in \cite{tachibana2018efficiently}), $L_{D}$ is the Critic's loss which should reach its maximum. The losses are unscaled.

\section{Experiments}

\subsection{Dataset design}

We experiment with VCTK 0.92v-mic2 \cite{vctk} and LibriTTS-500 \cite{zen2019libritts} datasets. We extract the feature set from silence-trimmed mono-channel waveforms with 22500 Hz sampling rate. For the feature extraction, we employ Praat \cite{boersma2001praat} with configuration as in \cite{babacan2013comparative} and for mel spectrograms (80-bins) we follow the setup in \cite{9054556}. We construct a smaller dataset than VCTK that includes speakers of the same gender by combining the female voices from VCTK and LibriTTS (VCTK+Libri-f). The exclusion of male voices improved the overall quality of the model. Due to the lack of singing training data, the model attempts to fill the missing pieces during synthesis with data from other speakers in the training set. This can result to a decrease in speaker similarity. However, following our suggested dataset design, this trait can be turned into an advantage and lead the model to a realistic result.

This intuition is reflected on the results of listening tests where 20 singing audio samples of different length, quality and gender are used as templates. Three different types of tests are conducted; Mean Opinion Score (MOS), Speaker Similarity (spk-sim), Song Similarity (song-sim) tests. On MOS test, 92 testers are asked to evaluate how natural the synthesized singing samples sound, rating the best sample with 5 and the worst with 1. The same scale is used for the other two tests too, each one involving 63 participants. They rate how similar the inferred samples sound either to a ground truth sample of the target speaker for spk-sim or to the original template track for song-sim. We present the results in Table \ref{table:femalemos}.

\begin{table}[htb]
	\caption{Mean Opinion Score (MOS), speaker similarity (spk-sim) and song similarity (song-sim) tests for VCTK target speakers s5 (20 min data), p276 (25 min data), p362 (5 min data).}
	\label{table:femalemos}
	\centering
	\begin{tabular}{llll}
		\toprule
	 &  \multicolumn{3}{c}{\textbf{Scores}}  \\
		\textbf{Dataset} &  \textbf{MOS} &  \textbf{spk-sim}  &  \textbf{song-sim}\\
		\midrule
		VCTK-f & 3.30 $\pm$ 1.13 & \textbf{3.34 $\pm$ 0.91}  & 3.35 $\pm$ 1.01 \\
		
		VCTK  & 3.32 $\pm$ 1.13 & 3.32 $\pm$ 0.93  & 3.38 $\pm$ 0.99\\
		
		VCTK+Libri-f  & \textbf{3.38 $\pm$ 1.08} & \textbf{3.45 $\pm$ 0.86} & \textbf{3.48 $\pm$ 0.93} \\	
		\bottomrule
	\end{tabular}
\end{table}

\subsection{Vocoder} 

The outputs of the acoustic model are passed to WaveGlow \cite{prenger2019waveglow} as used in \cite{9054556} without any interventions or modifications. We choose the WaveGlow vocoder for its stable performance over a large pitch range as also stated in \cite{perrotin21_interspeech}. The vocoder is trained on an internal dataset of a single female speaker, maintaining the singing-data-free setup end-to-end.

\subsection{Training Setup}

We train the acoustic model until 250k iterations with decoding step $r=5$ and the Critic from 150k to 250k iterations. After that point, the discrimination starts until 500k iterations. The starting learning rate is 0.0005 and is annealed to the half value every 100k iterations. All modules are trained with Adam optimizer, gradient clipping and batch size of 32. The training does not involve any data augmentation methods or pre-trained weights from other models; the input of model are only the phoneme sequence, the mel spectrogram and the individually normalized contours of voice features. 

\subsection{Inference and Cross-Lingual SVS}

During inference, our proposed feature set is extracted from the source audio of interest  that serves as a template for the voice synthesis. As in training, each contour of the feature set is individually normalized to be in accord with the feature value ranges of the target speaker in the training data. The model has some tolerance for mildly noisy input and allows control on the output through the feature encoder via relative percentage deviations from the input values in each dimension of the template. Textless inference takes place as a fallback solution for languages that are missing from the training set, allowing cross-lingual SVS. The most probable reason for this, is that linguistic information is derived from the formant information within the model rather than from the textual one. Early exploratory experiments on predicting formants from the text, made the model more responsive to zero text input during inference. 
The work in \cite{9362077} also proposes cross-lingual SVS with a different approach that also includes training speech data from the target language.
To our knowledge, Karaoker is the first application for textless cross-lingual SVS without any data or resources from the target language. 
The comparison and evaluation of cross-lingual and textless SVS is hard, therefore we provide some samples in our demo page for the reader to assess \footnote{ https://innoetics.github.io/publications/karaoker/index.html \label{demopage}}.

\subsection{Objective evaluation}

For the objective evaluation, we pick our best model (VCTK+LibriTTS-f) and we calculate: i) the root mean square error on the median-normalized pitches (\textbf{mF0 RMSE}) for pitch accuracy ii) the median cosine similarity of mean x-vectors \cite{snyder2018x} (\textbf{X-VEC COS}) for speaker similarity. We assess median-normalized pitch values because the outputs of the model are relative to the data coverage of the target speaker. We compute the metrics on the generated audio samples and the ground truth data (template audios for F0 / training data for x-vectors) for a random sample of 100 speakers in the training set. The results were \textbf{0.08 mF0 RMSE} and \textbf{68.7\% X-VEC COS}, suggesting that our model synthesizes singing voice following pitch templates accurately while achieving high target speaker similarity.

\subsection{Comparing with SOTA}

 Mellotron \cite{9054556} is a study that excludes the use of singing data in its pipeline for SVS, thus we choose to compare it with our model. We use the pre-trained weights on LJSpeech and code from the official repository and we infer samples with truncated samples from NUS-48-E dataset \cite{duan2013nus} as templates. We observe that Mellotron often produced unintelligible voice that followed the reference pitch. We also train Karaoker up to 250k (omitting GAN training phase) on LJSpeech data \cite{ljspeech17} in order to compare it with Mellotron  and use the same templates for inference. The model inferred singing voice with a bit degraded quality inherited by its training set but without the issues that Mellotron had. We encourage the reader to listen to the samples in our paper's demo page \footref{demopage}. 

 These facts convey that Karaoker presents an improvement for its topic. Karaoker performs better with a multi-speaker training set but even on a single speaker setting, it outperforms Mellotron, by actually generating singing voice. Our model is proved to be accurate and expressive in pitch due to its conditioning on continuous input in comparison with approaches like M-U \cite{choi2021melody} that rely on pitch quantization.

\subsection{Ablations}

To showcase how the suggested components contribute to the result, we train the the ablations listed in Table \ref{table:ablationmos} on VCTK+LibriTTS-female dataset. The baseline refers to a Tacotron 2 model with its standard mel and gate losses, which is conditioned with our suggested feature encoder. For inference, we select templates that yield results of fair quality to investigate how the model was guided to that output.

\begin{table}[htb]
	\caption{Results for the Mean Opinion Score (MOS) and song similarity (song-sim) tests (90 participants in total) for ablations trained on VCTK+LibriTTS-f and their generated samples for the VCTK target speaker s5.}
	\label{table:ablationmos}
	\centering
	\begin{tabular}{llll}
		\toprule
		\textbf{Ablation\{No\}} &  \textbf{MOS} & \textbf{song-sim} \\
		\midrule
		baseline\{1\}  & 3.58 $\pm$ 0.93 & 3.16 $\pm$ 1.13\\
		base + svd\{2\}  & 3.55 $\pm$ 1.05 & 3.17 $\pm$ 1.01 \\
		base + mel rate\{3\}  & 3.26 $\pm$ 1.09 & 2.92 $\pm$ 1.08\\
		base + svd / mel rate\{4\}  & 3.33 $\pm$ 1.10 & 2.97 $\pm$ 1.12 \\
		Karaoker without feature \\ decoders \& Critic\{5\}  & 3.42 $\pm$ 0.99 & 2.90 $\pm$ 1.03 \\
		
		Karaoker without mel \\ classifier \& Critic\{6\} & 3.52 $\pm$ 0.89 & 3.53 $\pm$ 0.95 \\
		
		Karaoker with loss scaling \\ \& without Critic\{7\}  & 3.58 $\pm$ 1.01 & 3.34 $\pm$ 1.07 \\
		Karaoker without Critic\{8\}  & 3.58 $\pm$ 0.89 & 3.50 $\pm$ 0.94 \\
		Karaoker\{9\}  &  \textbf{3.66 $\pm$ 0.95} &  \textbf{3.74  $\pm$ 0.82} \\
		\bottomrule
	\end{tabular}
\end{table}

In general, baseline ablations (baseline, basesvd, basemrate) have limited expressiveness and variation because they operate in the comfort zone of the training set's coverage. The rest of the ablations attempt to diverge from the immediate data coverage with unexpected side-effects and the listeners rated them lower as they most probably valued more the overall stability and quality than pitch accuracy. This fact highlights also how more difficult is to subjectively evaluate SVS models compared to TTS MOS evaluations. Ablation  \{7\} is an early attempt on scaling the multiple losses \cite{lin2021closer} of the model which did not bring any noticeable improvements in quality. The combination of each ablation (apart from \{7\}) leads to a synergy and this is demonstrated by the results for Karaoker (\{9\}) that also solidifies the presence of the GAN scheme in the model's training.

\section{Conclusions}

We have presented an acoustic model that can produce singing voice trained solely on speech training data without any specialized resources. The model can copy the style of a speaker/singer in the frame level with precision on pitch and speaker similarity. The samples are synthesized by a vocoder trained on speech data, maintaining the singing-data-free constraint end-to-end. The dataset design proved to be crucial for the performance of the model and we will consider in the future a design based on a speaker similarity criterion. Further work will focus on investigating more thoroughly the role of the formants in the model and how to allow control on lyrics during inference.  We are interested in employing an adaptive scaling scheme for the losses of the training objective and optimize the performance of the different components during training. Also, more experiments will take place in the architecture of the acoustic model but with the applied low-resource constraints of this study in mind.

\bibliographystyle{IEEEtran}

\bibliography{mybib}

\end{document}